# Improving Graphene-metal Contacts: Thermal Induced Polishing

Eliezer Fernando Oliveira[1,2], Ricardo Paupitz Santos[3], Pedro Alves da Silva Autreto[1,4], Stanislav Moshkalev[5], and Douglas Soares Galvão[1,2]

[1]Gleb Wataghin Institute of Physics, Universidade Estadual de Campinas, Campinas, SP, Brazil

[2]Center for Computational Engineering & Sciences (CCES), University of Campinas - UNICAMP, Campinas, SP, Brazil

[3]Institute of Geosciences and Exact Sciences, São Paulo State University (UNESP), Rio Claro, SP, Brazil

[4]Federal University of ABC, Center of Natural Human Science, Santo Andre, SP, Brazil

[5]Center for Semiconductor Components, State University of Campinas (UNICAMP), Campinas, SP, Brazil

ABSTRACT

*Graphene is a very promising material for nanoelectronics applications due to its unique and remarkable electronic and thermal properties. However, when deposited on metallic electrodes the overall thermal conductivity is significantly decreased. This phenomenon has been attributed to the mismatch between the interfaces and contact thermal resistance. Experimentally, one way to improve the graphene/metal contact is thorough high-temperature annealing, but the detailed mechanisms behind these processes remain unclear. In order to address these questions, we carried out fully atomistic reactive molecular dynamics simulations using the ReaxFF force field to investigate the interactions between multi-layer graphene and metallic electrodes (nickel) under (thermal) annealing. Our results show that the annealing induces an upward-downward movement of the graphene layers, causing a pile-driver-like effect over the metallic surface. This graphene induced movements cause a planarization (thermal polishing-like effect) of the metallic surface, which results in the increase of the effective graphene/metal contact area. This can also explain the experimentally observed improvements of the thermal and electric conductivities.*

**INTRODUCTION**

Graphene application in thermal management in electronic devices is very promising, since its thermal conductivity achieves values up to $5.3 \times 10^3$ W/mK [1]. However, good graphene/metal electrodes interfaces have been difficult to achieve. The structural graphene/metal mismatch drastically reduces thermal conductivities due to poor thermal contacts between the graphene and the metallic electrodes. Experimentally, one approach employed to address this issue has been high-temperature annealing [2].

When local annealing of a multi-layer graphene at high-temperature (T≥1000.0°C) is performed by a laser beam (Figure 1), it results in an improved thermal contact between graphene/metal interfaces; this could be achieved due to decreased rugosity of the metallic substrate. Despite that significant improvements in thermal conduction have been obtained in this way, the underlying mechanisms behind these processes are still elusive. In order to address this issue, we carried out fully atomistic reactive molecular dynamics simulations (FARMD). In particular, we investigated how the interface between graphene/metal changes during the annealing processes and how this influences the thermal properties.

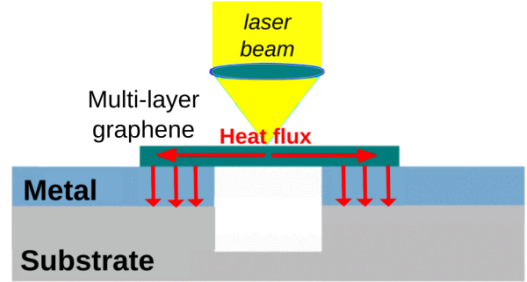

Figure 1: Layout of the experiment [2]. Graphene membranes are suspended over a metallic surfaces and laser beam irradiated.

## MATERIAL AND METHODS

All FARMD simulations were carried out using the ReaxFF force field [3], as available in the Large-scale Atomic/Molecular Massive Parallel Simulator (LAMMPS) code [4]. In order to mimic the experimental conditions [2], our structural models consist of passivated multi-layer (6 layers) graphene, each layer with dimensions of 36x26 Å$^2$ (334 atoms). The multi-layer structure is placed on top of a rugous nickel substrate (size of 64x52x40 Å$^3$, 5580 atoms); see Figure 2(a). Using a NVT ensemble, we performed an initial system thermalization, from 300.0 up to 1200.0 K for 1.0 ns and after this, another thermalization for more 1.0 ns at constant temperature of 1200.0K. In order to investigate whether the thermal properties improvement could be in part due to the decreasing of the metallic substrate rugosity, we built two systems (see the Figure 2(b): a passivated multi-layer graphene with 5 graphene layers (each layer with 42x80 Å$^2$ and 1404 atoms), which is placed on top of: i) two perfect nickel substrates (size of 64x118x26 Å$^3$ and 6118 atoms), and; ii) two nickel substrates similar to the i), but trenched (see Figures 2b and 3b). With these two systems, we performed an initial thermalization at 300.0K for 0.2 ns in a NVT ensemble. Then, in a NVE ensemble a heat flux was imposed from the Hot to the Cold regions (see the Figure 2(a)). The thermal conductivity K was calculated through Fourier law [5] (Equation 1):

$$K = \frac{1}{A}\frac{\Delta Q}{\Delta t}\frac{\Delta L}{\Delta T} \quad (1)$$

in which **A** is the cross-section area, **ΔQ/Δt** is the heat flux, and **ΔL/ΔT** is the temperature gradient along the heat flux direction.

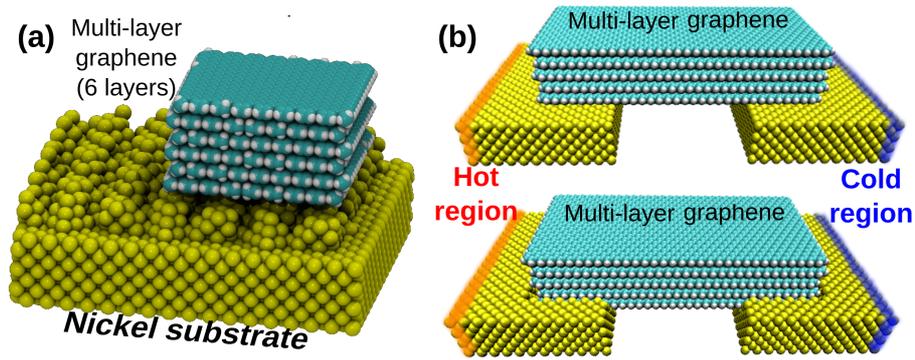

Figure 2: Schematic of the structural models used in the simulations. (a) graphene/metal during the annealing and; (b) simulation set up for the thermal transport calculations. Top: just deposited membranes; Bottom: 'trenched' ones (see also Figure 3b).

### RESULTS AND DISCUSSIONS

In Figure 3 we present MD snapshots of the system annealing at 1200K, where it is possible see the thermal induced polishing. Increasing the temperature, we observed from MD simulations trajectories an upward-downward movement of the graphene layers that increase as temperature increases. This graphene movement cause a pile-driver-like effect over the metallic surface and during the process, a planarization (polishing-like effect) of the metallic surface occurs. As we can see in Figure 3, besides the surface planarization, the graphene layers also enters/fuses into the metallic substrate, forming a valley. Thus, due to these processes, the graphene/metal contact area is increased during the thermal annealing, which can contribute to higher thermal conductivity.

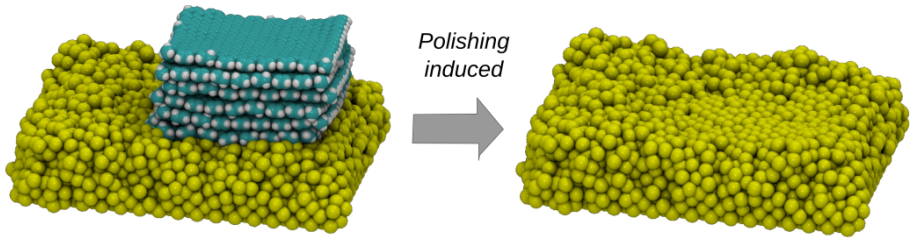

Figure 3: Resulting structure after the system annealing at 1200 K showing the thermal induced polishing effect caused by the multi-layer graphene movement.

In order to estimate these effects, we simulated the thermal conductivity of two configurations shown in Figure 2b. The temperature profiles for these systems when a heat flux is imposed are presented in Figure 4. As can be seen, the temperature difference between the metallic substrate and the multi-layer graphene is larger for the system with a flat metallic surface and occurs heat accumulation; this indicate that placing the multi-layer graphene into the valley promotes a better heat dispersion. Using the Fourier law (Equation 1), we can determine the thermal conductivities for both systems; in this way, we obtained a K=27.7 W/mK and K=78.8 W/mK for the systems with flat and trenched substrates, respectively. This indicates that the system with a trenched substrate is a better heat conductor. Using the obtained values for the thermal conductivities, we also can evaluate the thermal resistance [5] of both systems. The thermal resistance can be calculated by Equation 2:

$$R = \frac{Th}{K} \quad (2)$$

in which Th is the system thickness. In this way, we obtained a thermal resistance of $3.4 \times 10^{-10}$ m$^2$K/W and $0.7 \times 10^{-10}$ m$^2$K/W for the systems with flat and trenched substrates, respectively. These results show that the thermal annealing contributes to better thermal interfaces and heat conductivity into two steps: (i) an initial thermal induced surface polishing, which increases the contact area and (ii) if this process is maintained for longer periods and/or higher temperatures, the graphene layers 'fuses/sinks' into the metallic substrates further contributing the better thermal properties.

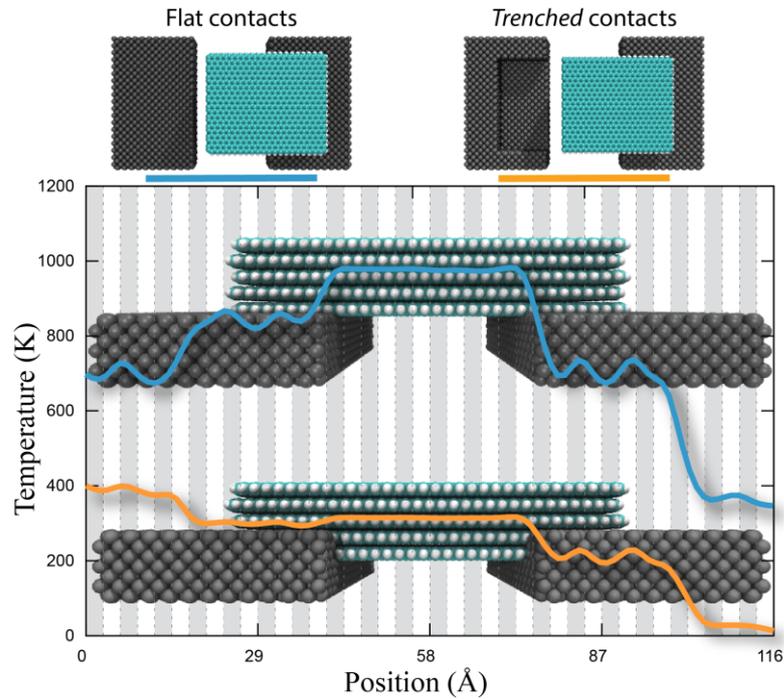

Figure 4: Temperature profiles when a heat flux is imposed for two systems in which the multi-layer graphene is placed on top of a flat and a trenched nickel substrate.

## CONCLUSIONS

We have investigated through fully atomistic reactive molecular dynamics simulations the thermal annealing of graphene/metal interfaces. From our MD simulations, we observed that due to the provided thermal energy the multi-layer graphene acquire an upward-downward movement that cause a thermal induced polishing-like (rugosity decrease) effect, which results in an increase of the contact area between graphene/metal. If the process is further continued the graphene 'fuses/sinks' into the metallic substrates, thus reducing the thermal resistance and also contribute to a better thermal graphene/metal interfaces.

## ACKNOWLEDGMENTS

We would like to thank the Brazilian agency FAPESP (Grants 2013/08293-7 and 2016/18499-0) for financial support and Center for Computational Engineering and Sciences (CCES) for the computational support.

## REFERENCES

1. A. A. Balandin, S. Ghosh, W. Bao, I. Calizo, D. Teweldebrhan, F. Miao and C. N. Lau *Nano Lett.* **8**, 902 (2008).
2. V. A. Ermakov, A. V. Alaferdov, A. R. Vaz, A. V. Baranov and S. A. Moshkalev, *Nanotechnology* **24**, 155301 (2013).
3. A. C. T. van Duin, S. Dasgupta, F. Lorant, and W. A. Goddard, *J. Phys. Chem. A* **105**, 9396 (2001).
4. S. J. Plimpton, *J. Comput. Phys.* **117**, 1 (1995).
5. Y. Shabany, *Heat Transfer: Thermal Management of Electronics*, CRC Press, Boca Raton, 2011.